\documentstyle[preprint,aps,epsf]{revtex}
\begin{document}
\draft

\title{Dynamic Vortex Phases and Pinning
in Superconductors with Twin Boundaries}
\author{C.~Reichhardt$^{1,2}$ C.J.~Olson$^{1,2}$, and Franco Nori$^1$}
\address{$1$. \ Department of Physics, The University of Michigan,
Ann Arbor, Michigan 48109-1120 \\
$2$. \ Department of Physics, University of California, Davis, CA 95616}

\date{\today}
\maketitle
\begin{abstract}
We investigate the pinning and driven dynamics of
vortices interacting with twin boundaries using
large scale molecular dynamics simulations on samples
with near one million pinning sites.
For low applied driving forces, the vortex lattice
orients itself parallel to the twin boundary and
we observe the creation of a flux gradient and vortex
free region near the edges of the twin boundary.
For increasing drive, we find evidence
for several distinct dynamical
flow phases which we characterize by the
density of defects in the vortex lattice,
the microscopic vortex flow patterns, and
orientation of the vortex lattice.
We show that these different dynamical phases can be
directly related to microscopically measurable
voltage--current $V(I)$ curves and voltage noise.
By conducting a series of simulations for various
twin boundary parameters we derive several vortex
dynamic phase diagrams.
\end{abstract}


\vskip1pc
\narrowtext

\newpage

\section{Introduction}

The understanding of vortex pinning and dynamics in high-Tc superconductors
is of great interest
for applications of superconductors  which require strong
pinning of vortices as well as the
rich variety of
behaviors that arise due to the competition
of a static or driven
elastic media with various forms of quenched disorder \cite{Blatter}.
The physics of a vortex lattice interacting with disorder is
relevant for
a wide variety of condensed matter systems including charge-density-waves,
driven Wigner crystals, magnetic bubble arrays, colloids,
Josephson junction arrays and superconducting wire networks, as well as
microscopic models of friction.

Twin boundaries are a very common defect found in
$YBa_{2}Cu_{3}O_{7-x}$(YBCO) and their
pinning properties have been extensively studied
using Bitter decoration \cite{Decorations},
torque magnetometry \cite{Gyorgy},
magnetization \cite{Schwartzendruber,Oussena,Lairson,Zhukov}
transport \cite{Safar},
magneto-optical imaging \cite{Duran,Vlaskov,Bishop,Welp,Olsson,rinke},
and theoretical studies \cite{review,theory,Groth}.
Many of the earlier experiments
on twinned  YBCO samples found conflicting evidence for the
role of twin boundaries in vortex pinning.
In particular, the
magneto-optical measurements by Duran {\it et al}. \cite{Duran}
had shown that twin
boundaries act as areas of reduced pinning that allow easy flux
penetration, whereas studies by Vlasko-Vlasov {\it et al}.
\cite{Vlaskov} found the twin boundaries to be barriers to 
flux motion. Further magneto-optical studies 
\cite{Bishop,Welp,Olsson,rinke}, systematic computer simulations 
\cite{Groth}, and transport measurements \cite{Safar}
have shown that these conflicting results can be resolved when
the direction of the Lorentz force with respect to the twin
boundary is considered. The twin boundary (TB)
acts as an easy-flow channel when the Lorentz force is
parallel to the twin, but acts as a strong barrier 
for forces perpendicular to the TB.

A very systematic simulational study, using samples with of the order of
a million pinning sites, by Groth {\it et al}. \cite{Groth} of the angular
dependence of the Lorentz force with respect to the twin boundary
showed that, when the angle between the Lorentz force and the twin
is large, a portion of the vortices get trapped inside the twin.
This produces a pile-up effect leading to a higher density
of vortices on one side of the twin in agreement
with observations by several groups including, for example, 
Vlasko-Vlasov {\it et al}. \cite{Vlaskov}, 
Welp {\it et al}. \cite{Welp}, and 
Wijngaarden {\it et al}. \cite{rinke}.
At lower angles between the Lorentz force and the twin, simulations
\cite{Groth} show that the flux moves in
channels along the twin boundary while some guided motion
of vortices along the edge of the  twin still occurs.
At the lowest angles
the flux flows most easily along the twin with a number of vortices
escaping from the twin and forming a flame pattern flux profile
in agreement with
magneto-optical experiments \cite{Duran,Vlaskov,Crabtree,rinke}.

Recently interest in vortex systems has strongly focused on
driven phases and dynamic phase transitions of vortices
interacting with random or periodic defects in superconductors.
The anisotropic pinning properties of twin boundaries as well as the
possibility of tuning the strength of the twin boundary pinning make
these defects quite distinct from random pinning or periodic pinning arrays,
so that new dynamical phases can be expected to appear.

In systems containing random pinning, experiments using
transport measurements \cite{danna-peak,Beasley,Andrei},
voltage noise measurements \cite{Rabin,danna},
vibrating reed measurements \cite{Zhang}, neutron scattering
\cite{Yaron}, and Bitter-decoration \cite{Pardo}, as well as
simulational work \cite{Koshelev,ShortDriven}, and work based
on perturbation and/or elasticity theory \cite{Giamarchi}
indicate that, at the depinning transition, the vortex lattice
may disorder and undergo {\it plastic flow\/} in which vortices
change nearest neighbors as mobile portions of the vortex
lattice tear past pinned portions.
At higher drives the vortex lattice may reorder and exhibit
elastic or ordered flow.  An intriguing question is whether
specific types of plastic flow exist, and how they could be
distinguished.  Simulations with randomly placed pinning
indicate the possible existence of at least two kinds of
plastic flow. The first type consists of well-defined
channels of mobile vortices flowing through the rest of
the pinned vortex lattice \cite{Brass,Jensen,Olson,Rivers}.
A second type consists of intermittent or avalanching motion
in which only a few vortices are mobile at any given time, but
over time all the vortices take part in the motion so that well
defined channels are not observed \cite{Jensen,Olson,Rivers}.

Recent simulations using the
time-dependent Ginzburg-Landau equations at $T = 0$ of vortices
interacting with twin boundaries have suggested the possibility of the
existence of three distinct flow phases which include two plastic flow phases
and an elastic flow phase \cite{Crabtree}.
Due to the nature of these simulations it was only possible
to consider three different driving currents for each
pinning parameter; so that $V(I)$ curves, voltage noise
signals, and the evolution of the vortex order
as a continuous function of increasing driving
force could not be extracted, nor could the evolution of the flow phases
with the system parameters be determined.

In order to examine the microscopic dynamics of vortices interacting with
twin boundaries we have performed large scale molecular dynamics
simulations for a wide variety of twin parameters which allow
us to carefully compare the different kinds of
plastic flow as a driving force is continuously increased.
Our results in this work complement our previous simulational
work on twin-boundaries \cite{Groth},
where we considered only the case of very slow driving that occurs as
a magnetic field is increased. In Ref.~\cite{Groth} we
considered flux-gradient-driven vortices and we focused on the
magnetic flux front profiles and compared them to magneto-optical
images. In this paper we focus on the microscopic aspects of
current-driven, as opposed to flux-gradient-driven, vortex motion
and structure as well as on transport measures.

\section{Simulation}

We consider
an infinite 2D slice in the $x$-$y$ plane of an
infinitely long (in the $z$ direction) parallelepiped. We use
periodic boundary conditions in the $x$-$y$ plane and simulate
stiff vortices that are perpendicular to the sample
(i.e, $ {\bf H} = H{\bf {\hat z}}$). These rigid flux lines can also
be thought of as representing the ``center of mass" positions of real,
somewhat flexible vortices, and the pinning in the bulk as representing
the average of the pinning along the length of the real vortex.
For flexible vortices, the bulk pinning can be on the same order
as the twin-boundary pinning even for large samples.
We numerically integrate the overdamped equations of motion:
\begin{equation}
{\bf f}_{i} = {\bf f}_{i}^{vv} + {\bf f}_{i}^{vp} + {\bf f}_{i}^{vTB} +
{\bf f}_{d} = \ \eta{\bf v}_{i} .
\end{equation}
Here, ${\bf f}_{i}$ is the total force on vortex $ i $, ${\bf f}_i^{vv}$ is
the force on the $i$th vortex from the other vortices,
${\bf f}_{i}^{vp}$ is the force from the vortex pin interaction,
${\bf f}_{i}^{vTB}$  is the force from the vortex-twin interaction, and
${\bf f}_{d} $ is the driving force; $ {\bf v}_{i}$ is the net velocity
of vortex $i$ and $ {\eta }$ is the viscosity, which is set equal to
unity in this work. The interaction between vortex $i$ and
other vortices is given by:
\begin{eqnarray}
\ {\bf f}_{i}^{vv} = \ \sum_{j=1}^{N_{v}} \ f_{0} \ K_{1}\left(
\frac{|{\bf r}_{i} - {\bf r}_{j}|}{ \lambda}\right) \
{\bf {\hat r}}_{ij} \ .
\end{eqnarray}
Here, ${\bf r}_{i}$ is the location of vortex $i$ and $ {\bf r}_{j}$ is the
location of vortex $j$, $f_{0} = \Phi_{0}^{2}/8\pi^{2}\lambda^{3} $,
$ \Phi_{0} = hc/2e$ is the elementary flux quantum, $ \lambda$ is the
penetration depth, $N_{v}$ is the number of vortices, and
$ {\bf {\hat r}}_{ij} = ({\bf r}_{i} - {\bf r}_{j})/|{\bf r}_{i} -
{\bf r}_{j}|$.
The force between vortices decreases exponentially at distances
greater than $ \lambda$, and we cut off this force for distances greater than
$6\lambda$. A cutoff is also placed
on the force for distances less than $ 0.1\lambda$
to avoid the logarithmic divergence of forces.
These cutoffs have been found to produce negligible effects for
the range of parameters we investigate here.
For convenience, throughout this work all lengths are measured in
units of $ \lambda$, forces in units of $f_{0}$, and fields in units of
$\Phi_0/\lambda^{2}$.

To model pinning in the bulk, we divide our system into a
$ 1000\times 1000$ grid where each grid element represents a pinning site. The
pinning density $n_{p}$ is $ 496/\lambda^{2}$, which is within
experimentally determined values. At each pinning site $(l,m)$ the
pinning force $f_{l,m}^{thr}$
is chosen from a uniform distribution $[0,f_{p}]$,
where $f_{p}$ is the maximum possible pinning force. If the
magnitude of the
force produced by the other vortices, driving force and twin boundaries
acting on a vortex located on a pinning site $(l,m)$ is
less than the threshold pinning force $f_{l,m}^{thr}$, the vortex
remains pinned at the pinning site. If the force on the vortex
is greater than $f_{l,m}^{thr}$, then the effective pinning force
$f_{i}^{vp}$
drops to zero and the vortex moves continuously until it encounters a pinning
site that has a threshold force greater than the net force on the
vortex. The pinning therefore acts as a stick-slip friction force with the
following properties
\begin{equation}
{\bf f}_{i}^{vp} \ =\ -\ {\bf f}_{i}^{net}, \ \ \ \ \ \ f_{i}^{net} < f_{l,m}^{thr}
\end{equation}
 and
\begin{equation}
{\bf f}_{i}^{vp} \ = \ 0, \ \ \ \ \ \ f_{i}^{net} > f_{l,m}^{thr}  \ \ .
\end{equation}

For the twin boundary pinning, we have considered a large number
of models, all giving similar results.  The simplest model that
is most consistent with experiments is that of an attractive well
containing stick-slip pinning with a different maximum threshold
force $f_{p}^{TB}$ than that of the bulk pinning outside the TB,
$ f_{p}$. This model of pinning is very similar to the one
inferred from the measurements in \cite{Oussena} where the TB
channel has strong depth variations.  The ratio $f_{p}^{TB}/f_{p}$
is expected to vary as a function of temperature.
In the case predicted for low $T$ \cite{Blatter}
where $f_{p}^{TB}/f_{p} < 1$,  the twin boundary acts
as an easy flow channel for certain angles \cite{Groth}.
On the other hand, at higher $T$,
$f_{p}^{TB}/f_{p} > 1$, and the twin acts as a barrier to
flux flow. This second case is the most similar to the simulations
conducted in \cite{Crabtree} where the twin boundary was modeled as
a line of parabolic pinning. In our simulations  we can mimic the
effects of temperature by varying the ratio of $f_{p}^{TB}/f_{p}$.

 The twin boundary itself is modeled as an attractive parabolic channel
with a width denoted by $ 2\xi^{TB}$. The force on the $i$th vortex due
to the $k$th the twin boundary is
\begin{eqnarray}
{\bf f}_{i}^{vTB} \; = \; f^{TB}
\left( \frac{ d_{ik}^{TB} }{ \xi^{TB} } \right) \
\Theta \! \left( \frac{ \xi^{TB} - d_{ik}^{TB} } { \lambda } \right)
\ {\hat {\bf r}}_{ik}
\end{eqnarray}
where $ d_{ik}^{TB}$ is the perpendicular distance between the $i$th vortex
and the $k$th twin boundary.

 The driving force representing the Lorentz force from an applied current
is modeled as a uniform force on all the vortices.
The driving force is applied in the
$x$-direction and is slowly increased linearly with time. We
examine the average force in the $x$-direction
\begin{equation}
V_{x} = \frac{1}{N_{v}}\sum_{i=1}^{N_{v}} \; {\bf v}_{i}\cdot{\bf {\hat x}} \ ,
\end{equation}
as well as the average force in the $y$-direction
\begin{equation}
V_{y} = \frac{1}{N_{v}}\sum_{i=1}^{N_{v}} \; {\bf v}_{i}\cdot{\bf {\hat y}} \ .
\end{equation}
These quantities are related to macroscopically measured voltage-current
{\it V(I)} curves.

We also measure the density of 6-fold coordinated vortices $P_{6}$.
Strong plastic flow causes an increase in the number of defects
and a corresponding drop in $P_{6}$, while
elastic flow is associated with few or no defects.
Another measure of order in the lattice is the average height of
the first-oder peaks in the structure
factor $S(k)$.
\begin{equation}
S({\bf k}) = \frac{1}{L^{2}}\sum_{i,j}e^{i{\bf k}
\cdot
({\bf r}_{i} -{\bf r}_{j})} \ .
\end{equation}

The defect density can
also be correlated with the voltage noise power spectra $S(\nu)$.
\begin{equation}
S(\nu) = \int V_{x}(t) \ e^{2\pi i\nu t} \ dt \ .
\end{equation}
A vortex lattice that is flowing plastically should produce a
large amount of voltage noise.  To measure the quantity of noise produced, we
integrated the noise power over one frequency octave \cite{Rabin,danna}.

\section{Dynamic Phases}

In order to directly observe the nature of the vortex flow in the
presence of twin boundaries, we have imaged the trajectories of the
moving vortices as the driving current along the $x$-axis is increased.
We find three types of vortex flow, which are shown in Fig.~1.
There, and for three different applied driving forces,
we show the vortex positions (dots) and the trajectories (lines)
that the vortices follow  when interacting with a
twin boundary (dotted line) that acts as a strong pinning
barrier for motion across the twin.
Here $f_{p} = 0.02f_{0}$, $f^{TB}_{p} = 1.0f_{0}$,
$f^{TB} = 0.15f_{0}$, with the twin boundary having
a width of $0.5\lambda$.  In Fig.~1(a) for
the lowest drive, $f_{d} = 0.05f_{0}$,
the vortex lattice is predominantly triangular, and
{\it aligned} with the twin plane.
The vortices that have struck the twin boundary
are pinned, while the remaining vortices
flow in  an orderly fashion at
a 45 degree angle from the $x$ axis,
as seen in Fig.~1(b).
The moving vortices do not {\it cross} the twin boundary
but are instead {\it guided} so that the vortices do not move
parallel to the
direction of the applied driving force. We term this phase
{\it guided plastic motion} (GPM), since vortex neighbors slip past
each other near the twin boundary. The vortices trapped
in the twin boundary
remain permanently pinned in this phase.
We also observe a build-up or a
higher density of flux lines along one side of the
twin boundary.
This type of density profile
has been previously observed in
flux-gradient driven simulations and magneto-optical experiments.

At higher drives, as shown in Fig.~1(c,d) with $f_{d} = 0.35f_{0}$,
there is a transition to a more disordered flow and the
vortices start to {\it cross} the twin boundary.  The
overall vortex structure [Fig.~1(c)]
is more disordered than it was at lower drives [Fig.~1(a)].
Unlike the guided plastic motion phase, the vortices pinned
along the
twin boundary are
only temporarily trapped, and occasionally escape from the twin and are
replaced by new vortices intermittently.
The vortex trajectories shown in Fig.~1(d) also indicate that
some vortex guiding still occurs. We label this phase the
{\it plastic motion} (PM) phase.
At even higher driving currents we observe
a transition from the plastic flow phase to an
{\it elastic motion} (EM) phase where the
effect of the twin boundary becomes minimal, as
shown in Fig.~1(e,f) for $f_{d} = 1.25f_{0}$.
Here, the vortex lattice reorders [Fig.~1(e)], the
vortices flow {\it along} the direction of the applied
Lorentz force [Fig.~1(f)], and no build-up of the flux
near the twin appears.

\section{Current-Voltage Characteristics and Vortex Structure}

In order to quantify the phases illustrated in Fig.~1,
we analyze the transverse $V_{y}$ and longitudinal $V_{x}$
average vortex velocities, as well as the six-fold
coordination number $P_{6}$ and the average
value of the first order peaks in the structure factor,
$<S(k)>$, as a function of applied drive.
As shown in Fig.~2,
for drives less then the bulk pinning,
$f_{d} < f_{p} = 0.02f_{0}$,
the vortex lattice is pinned and  $V_{y} = V_{x} = 0$.
For low drives, $0.02f_0 < f_{d} < 0.17f_{0}$,
the vortex velocities increase linearly with driving force, and
$V_{x} \approx V_{y}$ indicating that
the vortices are following the twin boundary by
moving at a $45^{\circ}$ angle, as was shown in Fig.~1(b).
The fraction of six-fold coordinated vortices, $P_{6} = 0.8$, 
remains roughly constant throughout the guided plastic motion
phase.  Above $f_{d}/f_{0} = 0.225$, two trends are observed.
First, the {\it longitudinal\/} velocity $V_{x}$ continues to
increase.  This trend can be better seen in the inset of Fig.~2(a),  
which has a larger range of values for the vertical axis in order
to monitor the linear growth over a wider range of velocities.
Second, the {\it transverse\/} velocity $V_{y}$
flattens and then begins to decrease, indicating that the
vortices have begun to move across the twin boundary.

The vortex lattice becomes slightly more disordered
in this plastic flow phase as indicated
by the drop in $P_{6}$ and the smaller drop in
$<S(k)>$. As $f_{d}$ is increased further, $V_{y}$
gradually decreases, but remains finite
as vortices cross the twin at an increasing rate.
When $V_{y}$ approaches zero, near $f_{d}/f_{0} = 0.85$,
the vortex lattice {\it reorders\/} as indicated by 
the increase in $P_{6}$ and $<S(k)>$.
We note that the reordering transition in $P_{6}$ is considerably
sharper than that typically observed in simulations with random
pinning.

\section{Noise Measurements}

An indirect experimental probe of the plastic vortex
motion is the voltage noise produced by the flowing flux.
During plastic flow the voltage noise is expected to be maximal.
Indeed, in simulations with random pinning \cite{ShortDriven,Rivers}, large
noise power was associated with the highly plastic motion of a
disordered vortex lattice. Further, large  noise is considered
a signature for plastic flow in the peak effect regime.
In order to compare the different plastic flow phases seen here with
those observed for random pinning, we measure the noise power for each
phase and plot the results in Fig.~3 along with the corresponding
$V_{y}$ versus $f_{d}$ curve.

The noise power is relatively low
in the GPM phase, increases to a large value in the PM phase,
and then gradually decreases as the EM phase is
approached.
In the GPM regime, although
tearing of the vortex lattice occurs
at the boundaries between the
pinned and flowing
vortices,
the vortex trajectories
follow fixed channels
and a large portion of the vortex lattice remains
ordered. This very orderly vortex motion produces little noise.
In the PM phase, the vortex lattice is highly disordered and the
trajectories follow continuously changing paths so the
corresponding voltage noise power is high.
This difference in noise power between the static and changing channels
for vortex flow agrees well with results obtained in systems
with strong random pinning.
In such systems, when
the vortex flow follows fixed winding channels that
do not change with time, low noise power is observed above the depinning
threshold \cite{ShortDriven,Rivers}.
Similarly, when the
pinning is weak and the vortices move in straight fixed
lines, low voltage noise is observed \cite{ShortDriven,Rivers}. This
latter case agrees well with the result seen here in the
GPM and EM phases, when
the vortices follow straight paths and produce little noise power.

\section{Dynamic Phase Diagrams}

To generalize our results to other parameters, we construct a phase
diagram of the dynamic phases.
We first measure the evolution of
$V_{x}$, $V_{y}$ and $P_{6}$
as a function of driving force for varying
$f_{p}^{TB}$ from $ 1.25f_{0}$ to $0.25f_{0}$.
These are seen in Fig.~4.  When the pinning strength $f_{p}^{TB}$
inside the twin increases, the width of the PM region grows, and
the amount of disorder in the vortex lattice increases,
as seen in the decrease of $P_{6}$.
 From the curves shown in Fig.~4, we construct a {\it dynamic 
phase diagram\/} which is plotted in Fig~5(a).
We determine the transition between the guided plastic and 
plastic flow phase from the onset of disorder in $P_{6}$ and the
downturn in $V_{y}$, whereas the plastic motion to elastic motion
transition line is marked at the point when $P_{6}$ begins to plateau.
The driving force $f_{d}$ at which
both the  GPM-PM and PM-EM transitions
occur each grow linearly with
$f_{p}$.  In particular, the PM-EM transition roughly follows
$f_{d} = f_{p}^{TB}$, indicating that
the vortex lattice  reorders
once the pinning forces are overcome.

It might be expected that
the transition out of the guided plastic motion phase
would fall at $f_{d} = f_{p}^{TB}$, when the vortices are
able to depin from the twin boundary.
Since vortex interactions are important, however, in actuality
the vortex density increases
on one side of the twin while a lower vortex density appears
on the other side. This localized flux gradient produces an additional
force on the vortices at the twin boundary, depinning them at
a driving force $f_{d} < f_{p}^{TB}$.
The additional force
from the flux-gradient is not spatially uniform,
unlike the driving force, so some of the vortices
will depin before others in a random manner. Once the applied driving
force and
the gradient force are large enough to start depinning vortices from
the twin,
the flux lines enter the plastic flow phase.
The effect of the pinning on the
vortices does not fully disappear until $f_{d} > f_{p}^{TB}$, however, which
is seen in the existence of a finite $V_{y}$. We also note that there
is a {\it pinned phase} where no vortex motion occurs
when $f_{d} < f_{p}$.

By changing the vortex density we can examine the effects of
changing the effective vortex-vortex interaction.
In Fig.~5(b) we plot the phase diagram
constructed from a series of simulations in which the
vortex density is varied.
As the vortex density decreases the GPM-PM and the
PM-EM transition lines shift to higher drives. This is 
because lower values of $B$ (or $N_v$) increase the effective
pinning force and shift the boundary to higher values of 
$f_d$.

\section{Conclusion}

We have examined the dynamics of driven superconducting vortices
interacting with twin boundary pinning.
We find three distinct flow phases as a function of driving force.
In the guided plastic motion phase,
the partially ordered vortex lattice flows in stationary
channels aligned with the twin boundary.
In this phase the transverse and longitudinal velocities
are equal and there is only a small amount of
noise in the velocity signals.
At higher drives, a flux gradient builds up along the twin
and the vortices begin to cross the twin boundary
intermittently. In this phase the vortex lattice
is disordered and a large amount of voltage noise appears.
The guiding effect of the twin gradually decreases for
increasing drives and the vortex lattice reorders,
producing an elastic flow phase.
By conducting a series of simulations
we have constructed phase diagrams both as a
function of twin boundary pinning strength
and as a function of the vortex density.
The phase boundaries all shift linearly in driving force as the
pinning strength increases. As the vortex density is lowered
the width of the guided motion region increases, while the
onset of the elastic motion phase is constant.

Twin boundaries correspond to one type of correlated pinning.
Another type involves periodic arrays of pinning sites \cite{paps}.
The dynamic phase diagrams of these structures with correlated 
pinning are also under current intense investigation.

{\it Note Added:} After completing this work we became aware
of the experiments in Ref.~\cite{Pastoriza}
which measure both the longitudinal and
transverse voltage signal for vortices driven in samples with
unidirectional
twin-boundaries. When the vortices are driven at 52 degrees with respect to
the twin boundaries, at low temperatures the vortex motion deviates
strongly from the direction of drive with a component moving along
the twin boundary. Using this experimental set-up it should be possible to
observe both the transverse and longitudinal vortex velocity as a function
of applied current.

\begin{figure}
\caption{
The vortex positions (left column) and flow patterns (right column) 
for three different applied drives.
Panels (a,b), with $f_{d}/f_{0} = 0.05$, show guided plastic motion.
Panels (c,d), with $f_{d}/f_{0} = 0.35$ show a slightly more disordered 
motion, exhibiting some plasticity, tearing, and healing.  
Specifically, the vortex lattice is slightly torn apart by the 
twin boundary, but it heals right after crossing it. 
Panels (e,f), with $f_{d}/f_{0} = 1.25$, show elastic flow.}
\label{fig1}
\end{figure}

\begin{figure}
\caption{
(a) The longitudinal $V_{x}$ and transverse $V_{y}$ average velocity versus
driving force for a system in which the twin boundary is represented by a
rough channel with strong pinning.  Here the twin has a maximum pinning
of $1.0 f_0$ and the point pinning has a maximum of $0.02 f_0$.  
 The inset of (a) shows how $V_{x}$ linearly
increases with $f_{d}$ for the same system in (a).
(b) The average six-fold coordination number $P_{6}$. (c) The
average magnitude $<S(k)>$ of the first order peaks of the
structure factor. In the disordered plastic motion regime
the low values of both $P_{6}$ and $<S(k)>$ reflect the 
large degree of disorder. The transition to the elastic 
flow regime is marked by the large increase in order 
indicated by $P_{6}$ and $<S(k)>$, as well as by the loss
of guided motion which occurs when $V_{y} \approx 0$.}
\label{fig2}
\end{figure}

\begin{figure}
\caption{
The average noise power $S_0$ versus driving force $f_{d}$ 
for the same system in Fig.~2.
In the guided plastic (low $f_d$) and elastic (high $f_d$) flow 
regimes the noise power is low.  However, in the 
intermediate-drive disordered plastic 
flow regime the noise power is high, and gradually decreases 
as $f_{d}$ is increased.}
\label{fig3}
\end{figure}

\begin{figure}
\caption{
(a) Transverse and longitudinal velocities versus driving force
for varying twin-boundary strengths: from top to bottom
$f_{p}/f_{0} = 0.25$, $0.5$, $0.75$, $1$, $1.25$.
The top curve corresponds to the case $f_{p}/f_{0} = 1.25$.
The second curve, from the top, corresponds to the same 
sample used in Figs.~1 and 2.
(b) The fraction of six-fold coordinated vortices 
$P_{6}$ versus driving. For increasing pinning strength
the width of both the guided and
plastic flow phases increases.}
\label{fig4}
\end{figure}

\begin{figure}
\caption{
(a) The dynamic phase diagram for the system in Fig.~4.
For increasing $f_{p}$ the PM-EM
transition line increases as $\propto f_{p}$ while the
GPM to PM transition increases much more slowly. (b) The dynamic
phase diagram for constant $f_{p}^{TB}$ but decreasing vortex density.
The PM-EM transition line remains roughly constant while the GPM-PM
transition line shifts to higher $f_{d}$ as the effective vortex-vortex
interaction decreases.}
\label{fig5}
\end{figure}

\setcounter{figure}{0}
\begin{figure}
\center{
\epsfxsize=5.4in
\epsfbox{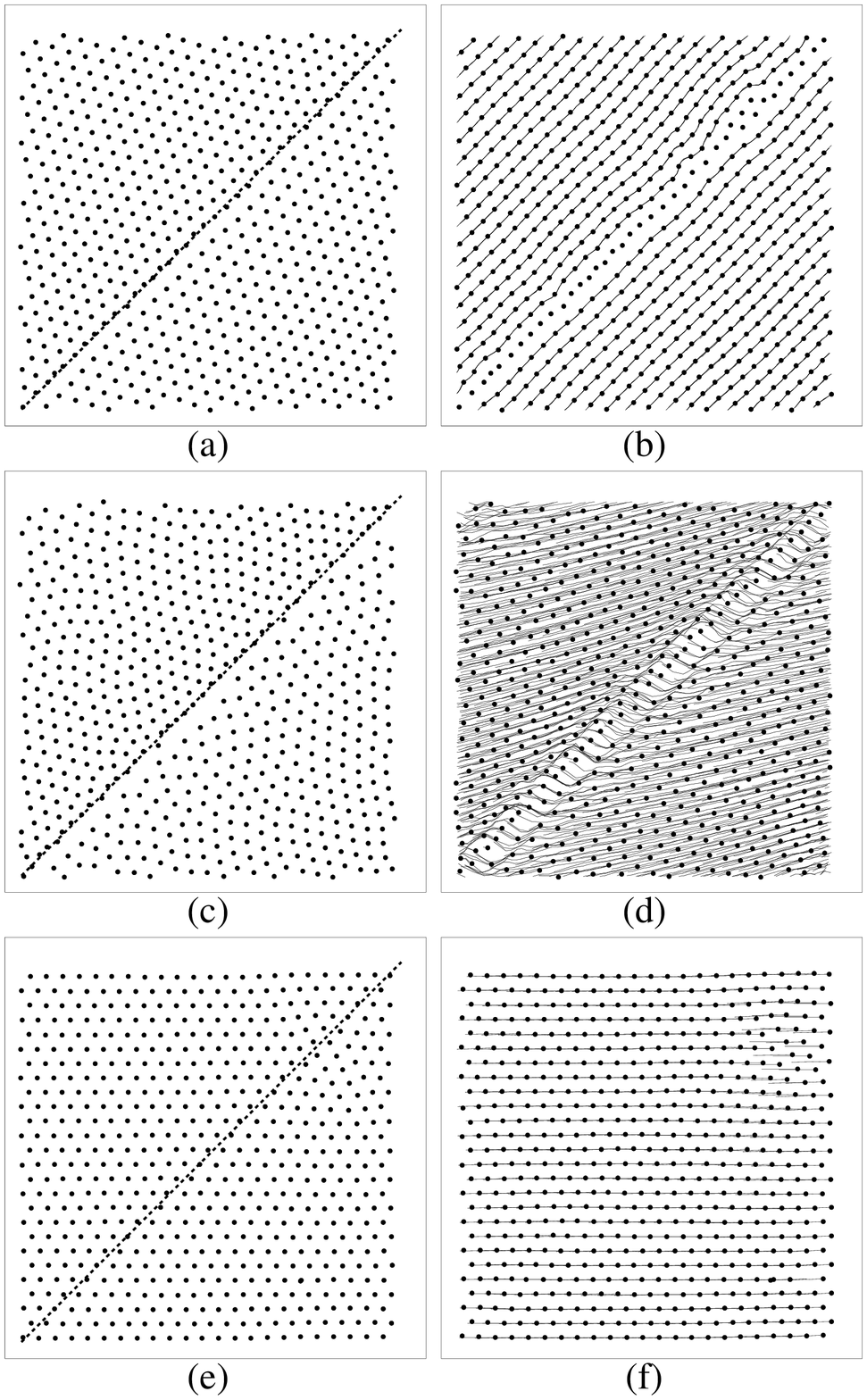}}
\caption{}
\end{figure}

\begin{figure}
\center{
\epsfxsize=6.0in
\epsfbox{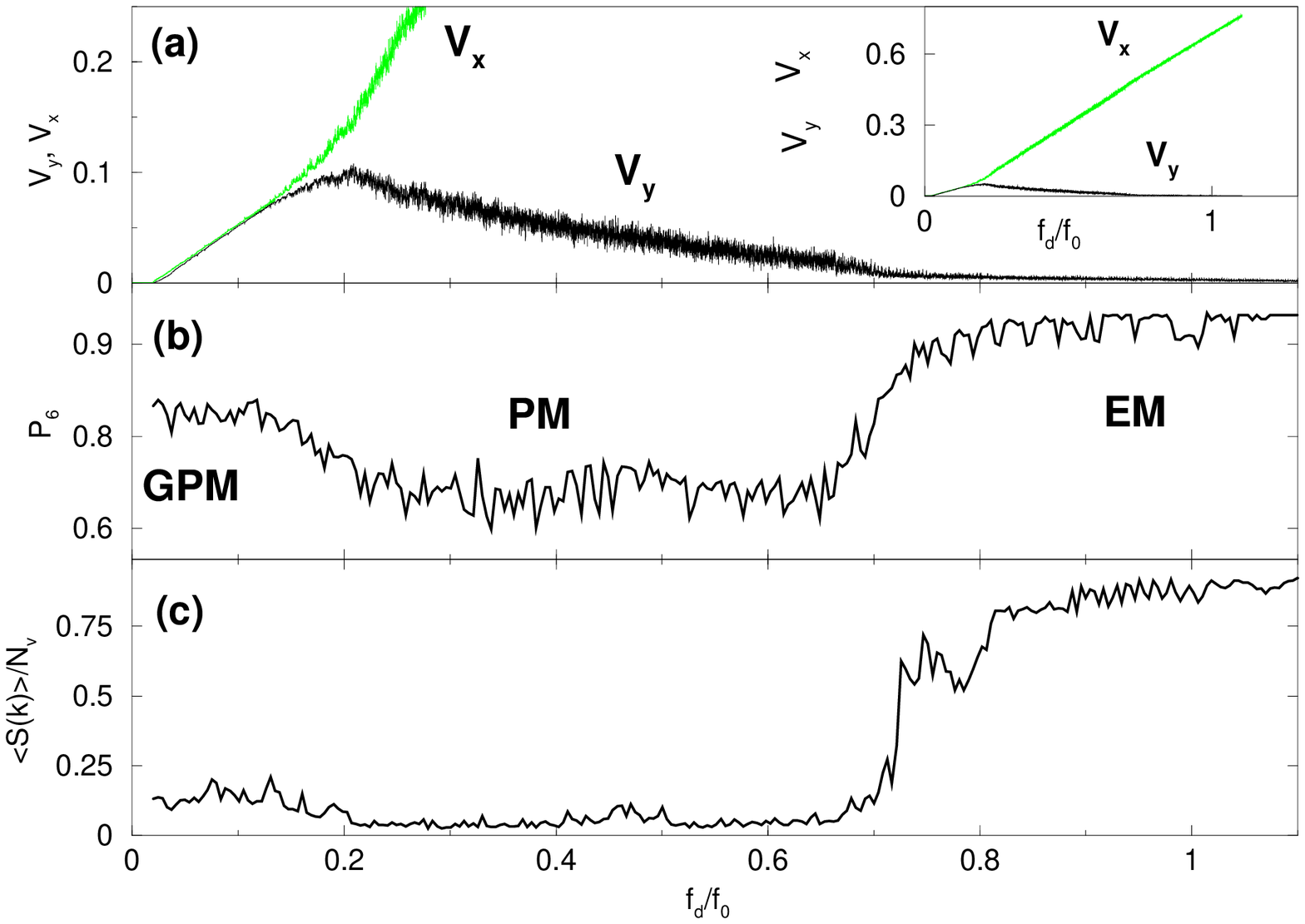}}
\caption{}
\end{figure}

\begin{figure}
\center{
\epsfxsize=6.0in
\epsfbox{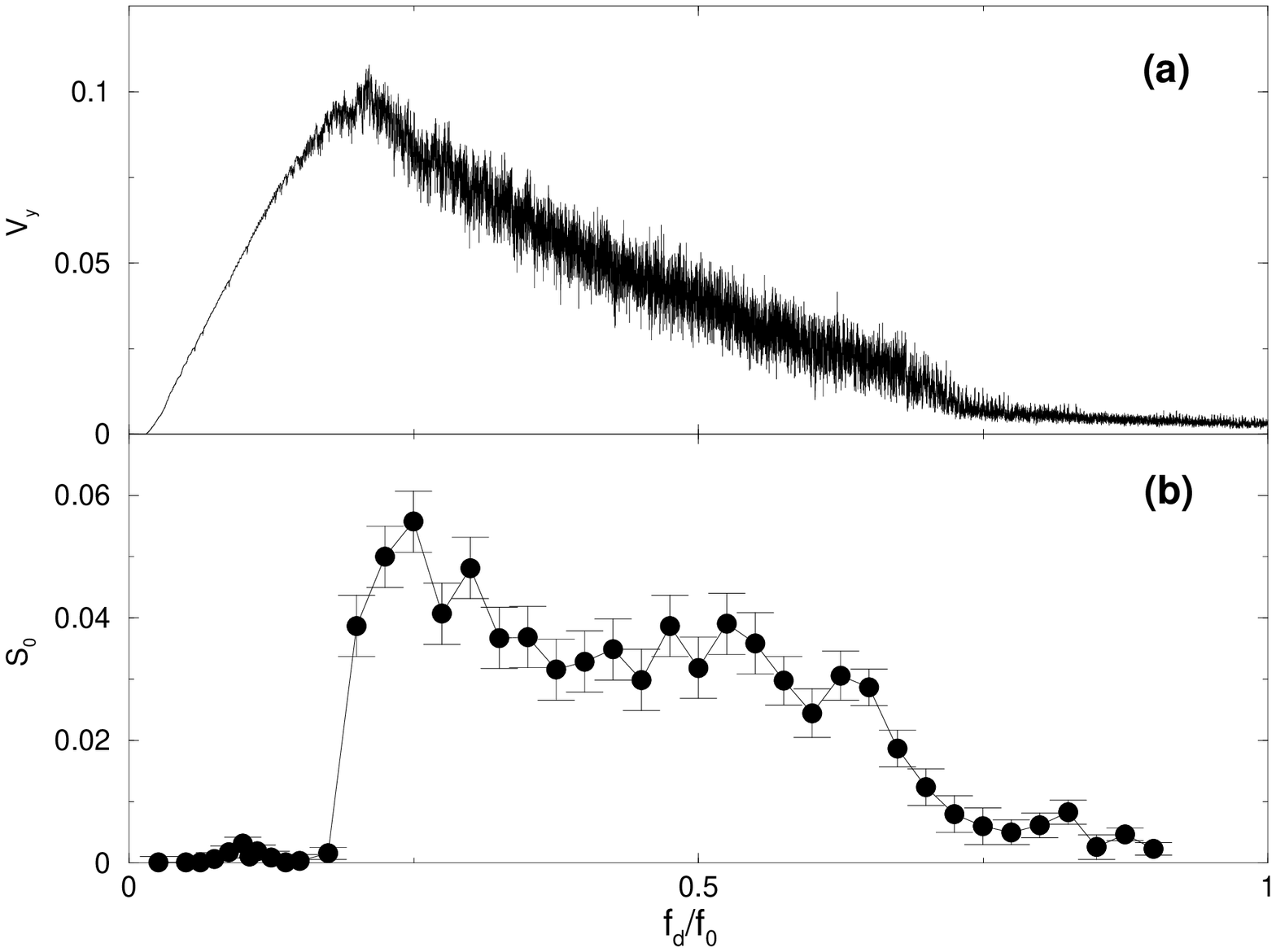}}
\caption{}
\end{figure}

\begin{figure}
\center{
\epsfxsize=6.0in
\epsfbox{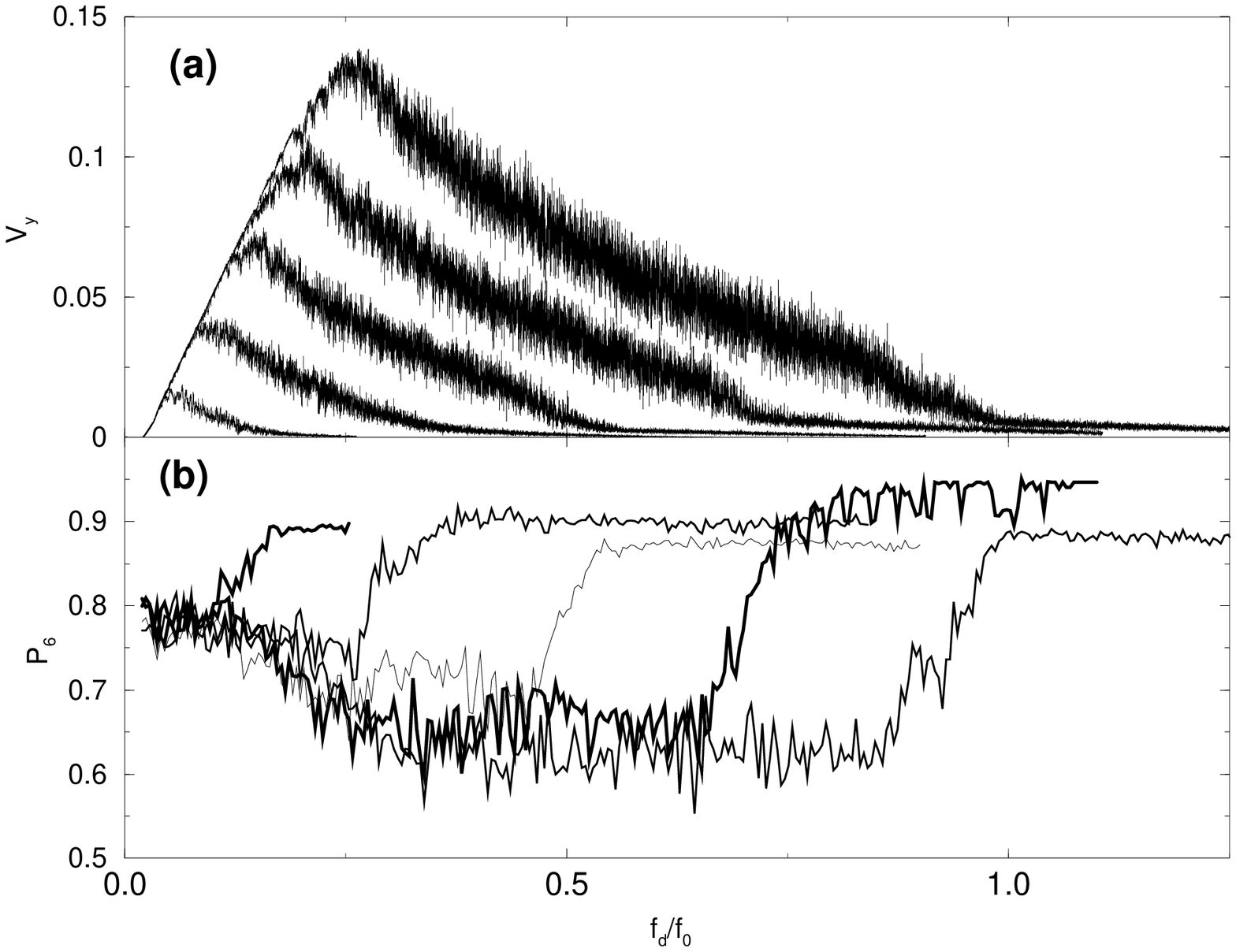}}
\caption{}
\end{figure}

\begin{figure}
\center{
\epsfxsize=6.0in
\epsfbox{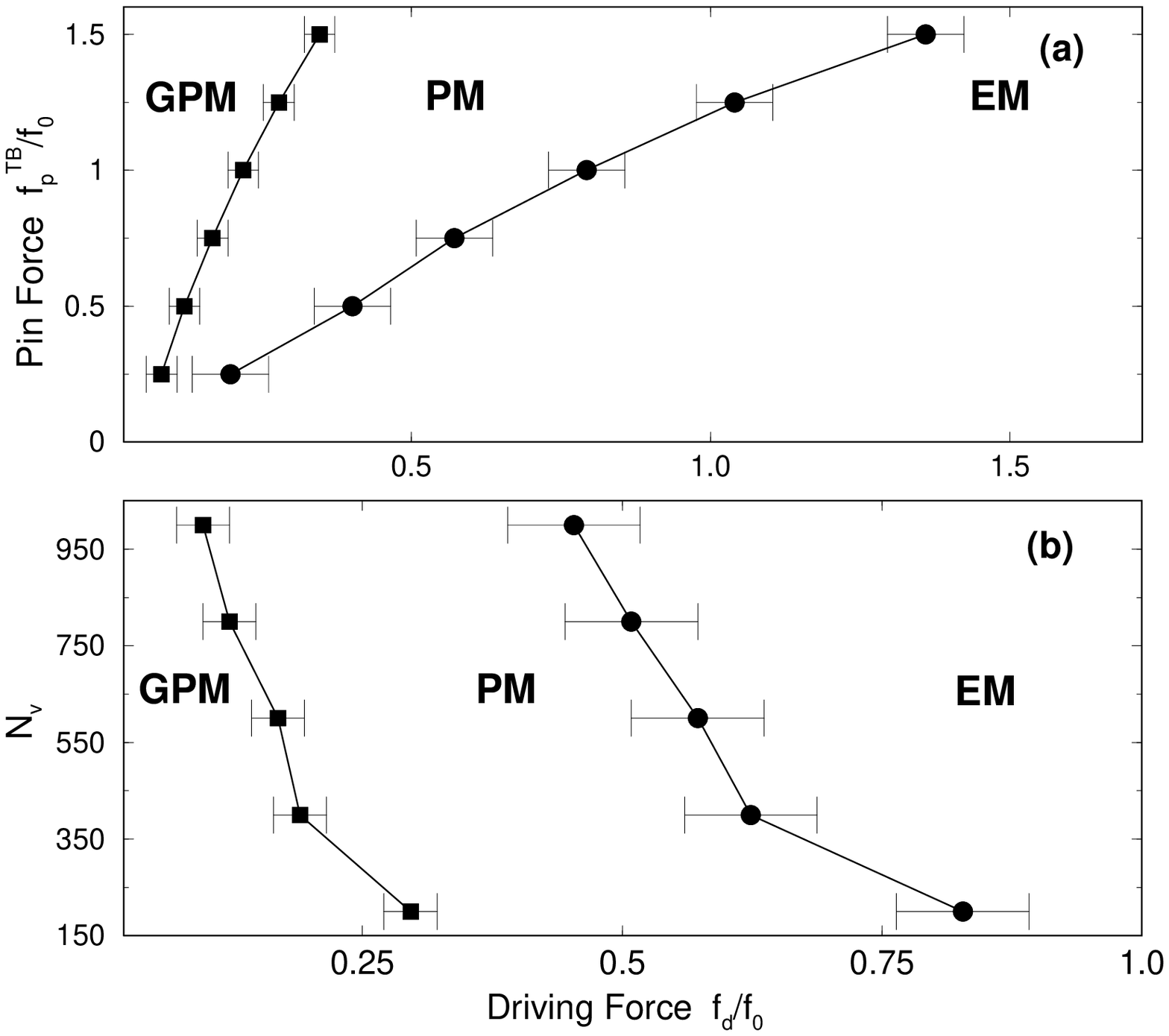}}
\caption{}
\end{figure}

\end{document}